\documentclass[letterpaper, 10 pt, conference]{ieeeconf}
\renewcommand{\baselinestretch}{0.98}

\usepackage{cite}
\usepackage{units}
\usepackage{amsthm,nccmath}
\newcounter{thm}
\newtheorem{prob}[thm]{Problem}

\usepackage{pgfplots}
\usepgfplotslibrary{groupplots}

\usepackage{hyperref}
\usepackage{amsmath}
\usepackage{amsfonts}
\usepackage{graphicx} % for pdf, bitmapped graphics files
\usepackage{amssymb}  % assumes amsmath package installed
\usepackage{pstricks,pst-plot,psfrag}
\usepackage{units}
\usepackage{float}
\usepackage{dsfont}
\usepackage{lettrine}
\usepackage{import}
\usepackage[acronym]{glossaries}

\usepackage{tikz}
\usetikzlibrary{backgrounds,calc,angles,quotes,arrows.meta,arrows,fit,shapes.geometric,shapes.multipart,pgfplots.groupplots, matrix,backgrounds}
\usetikzlibrary{circuits.ee.IEC}
\usetikzlibrary{shapes.gates.logic.US}
\usepackage{nth}
\usepackage{marvosym}%put packages in this document.
\newacronym{acr:cvt}{CVT}{continuously variable transmission}
\newacronym{acr:CoG}{CoG}{center of gravity}
\newacronym{acr:CoP}{CoP}{center of pressure}
\newacronym{acr:dp}{DP}{dynamic programming}
\newacronym{acr:DoF}{DoF}{degrees of freedom}
\newacronym{acr:ecms}{ECMS}{equivalent consumption minimization strategies}
\newacronym{acr:eltms}{ELTMS}{equivalent lap time minimization strategies}
\newacronym{acr:em}{EM}{electric motor}
\newacronym{acr:es2k}{ES2K}{Energy Storage to Kinetic}
\newacronym{acr:F1}{F1}{Formula 1}
\newacronym{acr:FIA}{FIA}{F\'{e}d\'{e}ration Internationale de l'Automobile}
\newacronym{acr:fgt}{FGT}{fixed-gear transmission}
\newacronym{acr:FD}{FD}{final drive}
\newacronym{acr:ice}{ICE}{internal combustion engine}
\newacronym{acr:k2es}{K2ES}{Kinetic to Energy Storage}
\newacronym{acr:mgu}{MGU}{motor generator unit}
\newacronym{acr:mguh}{MGU-H}{motor generator unit heat}
\newacronym{acr:mguk}{MGU-K}{motor generator unit kinetic}
\newacronym{acr:mpc}{MPC}{model predictive control}
\newacronym{acr:MS}{MS}{mini-sector}

\newacronym[description={energy management strategy}, \glslongpluralkey={energy management strategies},\glsshortpluralkey={EMSs}]{EMS}{EMS}{energy management strategy}%
\newacronym{acr:ODE}{ODE}{ordinary differential equation}
\newacronym{acr:pmp}{PMP}{Pontryagin's Minimum Principle}
\newacronym{acr:pu}{PU}{power unit}
\newacronym[description={powertrain operation}, \glslongpluralkey={powertrain operations},\glsshortpluralkey={POs}]{acr:PO}{PO}{powertrain operation}%
\newacronym{acr:socp}{SOCP}{second-order cone program}
\newacronym{acr:soe}{SoE}{state of energy}

\newcommand{\sN}{\mathbb{N}}

\newcommand{\sR}{\mathbb{R}}
\newcommand{\pushright}[1]{\ifmeasuring@#1\else\omit\hfill$\displaystyle#1$\fi\ignorespaces}
\newcommand{\pushleft}[1]{\ifmeasuring@#1\else\omit$\displaystyle#1$\hfill\fi\ignorespaces}
\makeatother
\newif\ifmargincomments %A quick way of turning off margin comments for, say, arXiv submission
\margincommentstrue
\newif\ifextendedversion %A quick way of turning off appendix
\ifmargincomments

\else

\fi
\usetikzlibrary{external}
\DeclareMathAlphabet{\mathpzc}{OT1}{pzc}{m}{it}
\IEEEoverridecommandlockouts

\newif\ifresponse
\responsetrue

\ifresponse

\else

\fi

\begin{document}

	\title{\bf Model Predictive Control Strategies for Electric Endurance Race Cars Accounting for Competitors’ Interactions
}

\author{Jorn van Kampen$^1$, Mauro Moriggi$^2$, Francesco Braghin$^2$ and Mauro Salazar$^1$
\thanks{$^1$Control Systems Technology section, Department of Mechanical Engineering, Eindhoven University of Technology (TU/e), The Netherlands.
		E-mails: {\tt\footnotesize \{j.h.e.v.kampen,m.r.u.salazar\}@tue.nl}
	} 
\thanks{$^2$Applied Mechanics sector, Department of Mechanical Engineering, Politecnico di Milano, Italy.
		E-mails: {\tt\footnotesize \{mauro.moriggi,francesco.braghin\}@polimi.it }
	}
}

\maketitle
\thispagestyle{empty}
\pagestyle{empty}

\begin{abstract}

This paper presents model predictive control strategies for battery electric endurance race cars accounting for interactions with the competitors.
In particular, we devise an optimization framework capturing the impact of the actions of the ego vehicle when interacting with competitors in a probabilistic fashion, jointly accounting for the optimal pit stop decision making, the charge times and the driving style in the course of the race.
We showcase our method for a simulated \unit[1]{h} endurance race at the Zandvoort circuit, using real-life data of internal combustion engine race cars from a previous event. 
Our results show that optimizing both the race strategy as well as the decision making during the race is very important, resulting in a significant \unit[21]{s} advantage over an always overtake approach, whilst revealing the competitiveness of e-race cars w.r.t.\ conventional ones.

%This paper presents an online race strategy optimization method that explicitly accounts for interactions with competitors to compute the optimal decision for a fully electric endurance race car. Thereby, we modify an existing race strategy optimization framework to capture the pit stop decision making per lap, and identify time penalties for every possible action to be taken by the ego vehicle. The online race strategy framework then optimizes the target lap times and energy budgets, together with the decision to charge the vehicle, from a lap-perspective for the current stint, and the future stint lengths and charge times from a stint-perspective. 
%Finally, we showcase our method for a simulated \unit[1]{h} race at the Zandvoort circuit, using real-life data of the competitors from a previous event. 
%Our results show that optimizing both the race strategy as well as the decision making during the race is very important, resulting in a significant \unit[21]{s} advantage over an always overtake approach, whilst revealing the competitiveness of e-race cars w.r.t.\ conventional ones.

% showcase 

\end{abstract}

%\begin{IEEEkeywords}
%	Electric vehicles, Endurance racing, Optimal control, Optimization, Race strategy
%\end{IEEEkeywords}

\section{Introduction}
The interest in the electrification of race cars has grown over the last 15 years, with the introduction both hybrid and fully electric racing classes~\cite{FE}. 
In the case of fully electric racing, it is of paramount important to make optimal use of the available battery energy---as it is the most limiting resource---strategically selecting charging stops and driving style through the course of a race.
This challenge gets an additional level of complexity once interactions with competitors are considered:
For instance, the overtaking of competitors are inevitable events that can require a significant amount of battery energy to successfully complete the maneuver. Moreover, competitors can also bring energy savings, as consumption can be reduced by driving in their slipstream, though at the cost of being restricted to their lap times.
These disturbances must be dealt with by carefully selecting the best action accounting for its impact on the overall race strategies and performance.
%These deviations in battery energy from a competitor-agnostic optimal solution, influence the optimal race strategy in terms of charge time and the number of laps to drive between charging stops (referred to as stint length). 
To address these challenges, this paper proposes model predictive control (MPC) algorithms that select the most appropriate actions when interacting with competitors, jointly optimizing the race strategies.

\subsubsection*{Related Literature}
This work pertains to two main research streams: race strategy optimization and simulations. 

Several authors have optimized the race from a lap perspective, for fully electric race cars~\cite{KampenHerrmannEtAl2023,Anselma2022b,LiuFotouhi2020} or hybrid-electric race cars~\cite{DuhrBuccheriEtAl2023}, either using an offline or adaptive online approach. These works were extended by accounting for competitor interactions in~\cite{PaparussoRianiEtAl2023}, whereby a set of overtaking strategies is defined offline and then used in an online framework employing Stochastic Dynamic Programming. Lastly, competitor decisions are predicted using a non-cooperative game approach~\cite{LinigerLygeros2020} or Gaussian Processes~\cite{ZhuBuschEtAl2023}. Yet these methods only capture the energy management strategy and not the impact on the pit stop strategy over the entire race. 

Race simulations including unexpected events and probabilistic effects are used to infer the optimal race strategy. These methods simulate entire races with tyre degradation, pit stops and overtake maneuvers~\cite{HeilmeierGrafEtAl2018}, and can potentially capture failures~\cite{BekkerLotz2009} or safety cars phases~\cite{HeilmeierGrafEtAl2020}. However, these methods choose the optimal strategy in terms of pit stop allocation and do not consider energy consumption. Furthermore, interactions are modeled as time losses, but no optimal decisions are taken regarding the interactions.  

In conclusion, to the best of the authors' knowledge, there are no control algorithms for electric race strategies including interactions with competitors. 
	
%	optimize both the high-level race strategy and the decisions during interactions in an online fashion, whilst accounting for the lap time, energy consumption and pit stop allocation.}{shrink}

%\begin{figure}[!t]
%	\centering
%	\input{./Figures/Framework_diagram.tex}                
%	\caption{Block diagram overview of the general approach to optimize the race strategy. The low-level framework provides the state trajectories and minimum stint time for several combinations of stint length and charge time. The high-level framework optimizes the race strategy for a given race time, using the stint time data.}
%	\label{fig:framework}
%\end{figure}
\begin{figure}
	\centering 
	\definecolor{mycolor1}{rgb}{0.00000,0.7500,1.0}%
\definecolor{mycolor2}{rgb}{1.000,0.65000,0.00}%
\definecolor{mycolor3}{rgb}{1.000,0.2000,0.200}%

\definecolor{teal}{rgb}{0.23,0.74,0.7}%
\definecolor{dark-teal}{rgb}{0,0.25,0.36}%
\definecolor{TUeRed}{rgb}{0.78,0.1,0.1}%

\tikzstyle{input} = [rectangle,rounded corners=2mm, minimum width=2cm, minimum height=0.5cm, text centered, draw=black, line width=0.25mm, fill=dark-teal,text=white]
\tikzstyle{output} = [rectangle,rounded corners=2mm, minimum width=2cm, minimum height=0.5cm,text centered, draw=black, line width=0.25mm, fill=teal,text=white]
\tikzstyle{Block} = [rectangle, minimum width=1cm, minimum height=0.5cm,text centered, draw=black, line width=0.25mm]
\tikzstyle{LUT} = [rectangle, minimum width=2cm, minimum height=0.5cm,text centered, draw=black, line width=0.25mm, fill=none,text=black, on background layer]

\tikzstyle{simpleNodeWide} = [rectangle, minimum width=1.5cm, minimum height=1cm,text centered, draw=black, line width=0.25mm]
\tikzstyle{arrow} = [->,-triangle 45,line width=0.25mm,black]
\tikzstyle{lineE} = [-,line width=0.5mm,gray!50,solid]
\tikzstyle{lineM} = [-,line width=0.25mm,black]
\tikzset{
	relative at/.style n args = {3}{
		at = {({$(#1.west)!#2!(#1.east)$} |- {$(#1.south)!#3!(#1.north)$})}
	}
}

\begin{tikzpicture}[node distance=0.7cm]
	\tikzstyle{every node}=[font=\scriptsize] % Font size
	\scriptsize
	\coordinate (Zero) at (0,0);
%	\node (Maps) [Block, align=center] at (Zero) {Lap time maps};
	
	\node[ inner sep=0pt,fill=white] (Maps) at (Zero) % distance of low level to lookup
	{\resizebox{2.5cm}{!} {\input{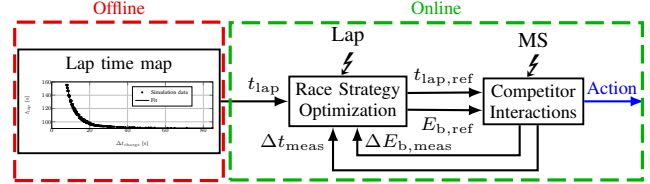}}};
	\node[ align=center, text=black,inner ysep=0] (LUTtext) at ($(Maps.north)+(90:0.2cm)$) {Lap time map}; % Should be on background
	
	\begin{scope}[on background layer]
		\node [LUT,fit=(Maps) (LUTtext)] (LUTbox) {};
	\end{scope}
	
		\node (MPC) [Block, align=center,anchor=west] at ($(LUTbox.east) + (0:0.9cm)$) { Race Strategy \\  Optimization};
			\node (Int) [Block, align=center,anchor=west] at ($(MPC.east) + (0:1cm)$) {Competitor \\ Interactions};
%				\node (Car) [Block, align=center] at ($(Int.east) + (0:1cm)$) {Vehicle};
	
%	
%	
%	
%	\node (Isl) [input, align=center] at ($(Ict) + (90:0.75cm)$) {Stint length};
%	\node (Irt) [input, align=center] at ($(Isl) + (90:3.3cm)$) {Race time}; % Distance of low level and high level
%	\node (HLP) [OCP, align=center] at ($(Irt) + (0:3cm)$) {High-level \\ optimization};
%	\node (Osl) [output, align=center] at ($(HLP) + (0:3cm)$) {Stint lengths};
%	\node (Ops) [output, align=center, above of=Osl] {Pit stops};
%	\node (Oct) [output, align=center, below of=Osl] {Charge times};
%	\node (LLP) [OCP, align=center] at ($(Ict) + (0:3cm)$) {Low-level \\ optimization};
%	\node (Ovs) [output, align=center]  at ($(LLP) + (0:3cm)$){Vehicle state \\ trajectories};
%	

	\node[anchor=south,inner sep=0pt,label=above :{\footnotesize Lap}] at (MPC.north) {\large\Lightning } ;
	\node[anchor=south,inner sep=0pt,label=above :{\footnotesize MS}] at (Int.north) {\large\Lightning } ;
	\draw [-latex,thick] (LUTbox.east) -- (MPC.west) node[above left] {$t_\mathrm{lap}$};  
	\draw [-latex,thick] (MPC.8) -- (Int.170.5) node[above left] {$t_\mathrm{lap,ref}$};
	\draw [-latex,thick] (MPC.-8) -- (Int.-170.5) node[below left] {$E_\mathrm{b,ref}$};
%	\draw [-latex,thick] (Int.east) -- (Car.west);
	\draw [-latex,thick] (Int.-80) --++ (-90:0.6cm) -| (MPC.-120) node[below left] {$\Delta t_\mathrm{meas}$};
	\draw [-latex,thick] (Int.-120) --++ (-90:0.4cm) -| (MPC.-70) node[below right] {$\Delta E_\mathrm{b,meas}$};
	\draw [-latex,thick,blue] (Int.east) --++ (0:0.8cm) node[above left,text=blue] {Action};
%	\draw [-latex,thick] (HLP.east) -- (Oct.west);
%	
%	\draw [-latex,thick] (LLP.east) -- (Ovs.west);
%	\draw [-latex,thick] (Ict.east) -- (LLP.west);
%	\draw [-latex,thick] (Isl.east) -++(0.5,0) |- (LLP.west);
%	
%	
%	
%	\draw [-latex,thick] (LLP.north) -- (LUTbox.south);
%	\draw [-latex,thick] (LUTbox.north) -- (HLP.south);
%	
	\node (BoxOn) [rectangle,dash pattern=on 7pt off 3pt, draw=black!30!green, very thick, fit = (MPC) (Int),inner xsep=22pt,inner ysep=20pt ] {};
	%	\node (split) [rectangle,dashed, draw=black, line width=0.5mm, fit = (Ict) (Isl) (Ovs) (lookup)] {};
	%	\draw [dashed] ($Box.west|-($(lookup.north)!0.5!(Oct.south)$))--(Box.east,($(lookup.north)!0.5!(Oct.south)$));
%	\coordinate (xsplit) at ($(LUTbox.west)!0.5!(Oct.south)$);
	\node (BoxOff) [rectangle,dash pattern=on 7pt off 3pt, draw=black!10!red, very thick, fit = (LUTbox), inner ysep=10pt,inner xsep=1pt  ] {};
	\node[above,text=black!10!red] at (BoxOff.north) {Offline};
	\node[above,text=black!30!green] at (BoxOn.north) {Online};

	%	\node (INV) [simpleNode, align=center, right of=EM] {INV};
	%	\node (BAT) [simpleNodeWide, align=center, right of=INV] {};
	%	\node[relative at={BAT}{0.5}{0.25}] {BAT};
	%	\node[relative at={BAT}{0.35}{0.75}] {\small{$E_\mathrm{b}$, $\vartheta_\mathrm{b}$}};
	%	\node[relative at={EM}{0.5}{0.25}] {EM};
	%	\node[relative at={EM}{0.30}{0.75}] {\small{$\vartheta_\mathrm{m}$}};
	%	\node (Wheel) [simpleNode, align=center, below of=FD, yshift=1cm, minimum height=0.5cm, draw=none, fill=gray!50, rounded corners] {Wheel};
	%	%%% Edges %%%
	%	\draw [lineM] ($(BAT.west)+(0,-0.075)$) to [out=0,in=270] ($(BAT.north)+(0.50,0)$);
	%	\draw [lineM] ($(EM.west)+(0,-0.075)$) to [out=0,in=270] ($(EM.north)+(0.25,0)$);
	%	\draw [lineE] (FD.south) -- node[right] {} ($(Wheel.north)-(0,0.1)$);
	%	\draw [arrow] (Wheel.west) -- node[above, yshift=0.1cm] {$P_\mathrm{p}$} ($(Wheel.west) + (-0.75,0)$);
	%	%\draw [arrow] (Wheel.east) -- node[above, yshift=0.1cm] {$P_{\mathrm{brake},i}$} ($(Wheel.east) + (0.75,0)$);
	%	\draw [arrow] (EM.west) -- node[above, yshift=0.1cm] {$P_\mathrm{m}$} (FD.east);
	%	\draw [arrow] (INV.west) -- node[above, yshift=0.1cm] {$P_\mathrm{ac}$} (EM.east);
	%	\draw [arrow] (BAT.west) -- node[above, yshift=0.1cm] {$P_\mathrm{dc}$} (INV.east);
	%	\draw [arrow] (BAT.south) -- node[right, xshift=0.1cm] {$P_\mathrm{aux}$} ($(BAT.south) - (0,0.75)$);
\end{tikzpicture}
	\caption{Overview of the controller architecture. The race strategy optimization uses lap time maps to compute reference trajectories every lap, while the decision making algorithm provides an action and feeds back the actual states.}\vspace{-15pt}
	\label{fig:control_scheme}
\end{figure}

\subsubsection*{Statement of Contributions}
This paper presents MPC algorithms that optimize electric race strategies together with competitors' interactions in an online fashion.
The online adaptive race strategy optimization framework computes the race strategy in terms of target lap times, energy consumption and pit stop strategy, at the beginning of every lap.
When an interaction with a competitor is predicted to occur, the decision-making process is started to find the action that results in the lowest time penalty w.r.t.\ the original strategy. A schematic overview of the controller is shown in Fig.~\ref{fig:control_scheme}. Finally, we showcase our framework for a \unit[1]{h} race on the Zandvoort circuit with a fully electric vehicle competing against \gls{acr:ice} race cars.

% mention we compete against ICE cars

\subsubsection*{Organization}
The remainder of this paper is structured as follows: Section~\ref{sec:MPC} presents the online race strategy optimization framework, after which Section~\ref{sec:competitor interaction} describes the action selection process. We showcase our framework in Section~\ref{Results} and draw conclusions together with an outlook in Section~\ref{Conclusion}.

\section{Online Race Strategy Optimization} \label{sec:MPC}
% Lap time maps? Does not need a separate section. Describe the delta t charge in the MPC section.
% MPC
% - Lap and stint based
% - Full race horizon
% - delta t charge instead of energy based
% - lap maps with delta t charge i.o. dE_b (from pre-computing). This way, we can directly get the charge time pentaly.
% - If pit, reset cumulative charge time
% last stint? 
% Lap times from stint? -> interpolate or skip
% First lap is flying lap due to rolling start
% Lap map generated using previous framework
%

This section presents a model predictive controller (MPC) to optimize the race strategies online.
Specifically, we devise the framework inspired by\cite{KampenHerrmannEtAl2023}, extending it to optimize the current stint on a lap basis instead of stint basis, and implement it online in a receding horizon fashion.
The goal is to choose the optimal stint lengths (being the number of laps in between two pit stops), charge times and number of charging stops over the course of the entire race and adapt them online at every lap.
We set the elapsed time as the main state variable and convert charged energy to an equivalent charge time.

In endurance racing, we are typically aiming at maximizing the driven distance $S_{\mathrm{race}}$ within a certain total race time $T_\mathrm{race}$.
Optimizing the race strategy online, we need to decide the amount of energy to spend per lap and whether to make a pit stop to charge the car on a lap basis.
Modeling an entire endurance race on a lap basis would result in a computationally taxing problem due to the presence of several binary decision variables proportional to the number of laps.
Against this issue, one could optimize the race in a moving horizon fashion. Yet such an approach would potentially introduce sub-optimal results, as in an endurance race it is important to account for the end of the race when planning the strategies~\cite{KampenHerrmannEtAl2023}.
Therefore, we devise a hybrid formulation whereby we optimize the current stint on a lap basis and the rest of the race on a stint basis.
In this context, the total driven distance can then be formulated as
\par\nobreak
\begingroup
\allowdisplaybreaks
\begin{small}
	\begin{equation}
		\max S_{\mathrm{race}} = \max \sum_{i=1}^{n_{\mathrm{stops}}} S_{\mathrm{lap}}\cdot N_{\mathrm{laps}}(i) + \sum_{j=1}^{n_{\mathrm{laps}}} S_{\mathrm{lap}}\cdot b_{\mathrm{lap}}(j),
	\end{equation}
\end{small}%
\endgroup
where $n_{\mathrm{stops}}$ is the pre-defined number of pit stops, $N_{\mathrm{laps}}(i)\in\sN, \  \forall~i \in \left[1,...,~n_{\mathrm{stops}}~-~1\right]$ is the stint length with $\sN$ the set of natural numbers, and $S_{\mathrm{lap}}$ is the length of one lap. Similarly, we have $n_{\mathrm{laps}}$ as the pre-defined number of laps in the current stint, and $b_{\mathrm{lap}}(j)$ is a binary variable representing whether lap $j$ is driven. Since the vehicle starts and stops at the pit box, the stint length should be an integer number of laps. As it is unlikely that the vehicle is exactly at the finish line when the race time limit is reached, we allow the final stint length to be a non-integer number of laps, i.e., $N_{\mathrm{laps}}(n_\mathrm{stops})\in \sR_+$.  

\subsection{Current Stint}
\begin{figure}
	\vspace{4pt}
	\centering 
	\input{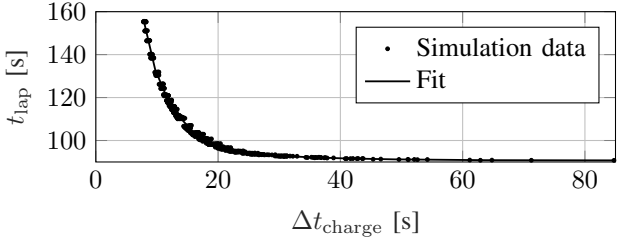}\vspace{-5pt}
	\caption{Lap time map for the base-lap using $n_\mathrm{fits}=50$ piecewise affine functions. The normalized root-mean-square error (RMSE) of the model is 0.012\% w.r.t. the maximum lap time.}\vspace{-15pt}
	\label{fig:fit_lap_map}
\end{figure}

As the current stint is modeled on a lap basis, we make use of pre-computed lap time maps, generated using the framework in~\cite{KampenHerrmannEtAl2023}, that correspond to the minimum achievable lap time for a given energy budget per lap. However, since the main state is defined as a charge time rather than a charged energy, we reformulate the energy budget per lap to an equivalent charge time using a pre-defined charging profile. The lap time maps are then approximated using the piecewise affine and convex functions as
\par\nobreak
\begingroup
\allowdisplaybreaks
\begin{small}
	\begin{equation}
			t_{\mathrm{lap},k} =\max_{n\in [1,\dots, n_\mathrm{fits}]}\{c_{\mathrm{1},k,n}\cdot \Delta t_{\mathrm{charge}} + c_{\mathrm{0},k,n}\}, \  \forall n\in [1,\dots, n_\mathrm{fits}],
	\end{equation}
\end{small}%
\endgroup   
where $n_\mathrm{fits}$ is the number of piecewise affine functions used, $t_{\mathrm{lap},k}$ is the obtained lap time, with $k\in \{\text{base,in}\}$, $\Delta t_{\mathrm{charge}}$ is the equivalent charge time corresponding to the energy used per lap and $c_{\mathrm{1},k,n}, c_{\mathrm{0},k,n}$ are fitting parameters. Note that we distinguish between normal base-laps and in-laps, whereby the latter includes driving into the pit lane to charge the car. Out-lap maps are not included, since they are always part of a future stint, removing the need to model them on a lap basis. Fig.~\ref{fig:fit_lap_map} shows the lap time map for a base lap.
We observe that the slight misalignment of the data with the affine fit is due to the dependency of both the charge power and battery efficiency on the battery energy.
%\msmargin{To further improve the lap time model, one could consider adding the battery energy as a third dimension, but this is considered to be outside of the scope of this paper as the model is considered to be sufficiently accurate. }{consider deleting if you need space}

% x acknowledge that fit is not perfect, but outside of scope for now.
% x mention in lap and base lap
% x Show figure of fit of base and report RMSE

The total time to complete the remainder of the current stint $t_\mathrm{stint,current}$ is defined as the sum of all lap times
\par\nobreak\vspace{-5pt}
\begingroup
\allowdisplaybreaks
\begin{small}
	\begin{equation}
		t_\mathrm{stint,current} = \sum_{j=1}^{n_{\mathrm{laps}}-1} (t_\mathrm{lap,base}(j)) + t_\mathrm{lap,in}. \label{eq:t_stint_current}
	\end{equation}
\end{small}%
\endgroup  
where we constrain the base lap time as
\par\nobreak\vspace{-5pt}
\begingroup
\allowdisplaybreaks
\begin{small}
	\begin{multline}
		t_{\mathrm{lap,base}}(j) \geq c_{\mathrm{1,base},n}\cdot \Delta t_{\mathrm{charge}}(j) + c_{\mathrm{0,base},n}\\
		 - M\cdot(1-b_\mathrm{lap}(j)), \quad \forall j\in [1,\dots, n_\mathrm{laps}-1], \label{eq:t_lap}
	\end{multline}
\end{small}%
\endgroup  
with $M\gg t_\mathrm{lap,base,max}$. This constraint will then hold with equality at the optimum, since it is optimal to minimize lap time. We constrain the in-lap in a similar fashion, but using $j=n_\mathrm{laps}$, since the in-lap is always the final lap of the current stint. By introducing the binary variable $b_\mathrm{lap}$ and constraining the lap time to be non-negative
\par\nobreak\vspace{-5pt}
\begingroup
\allowdisplaybreaks
\begin{small}
	\begin{equation}
		t_{\mathrm{lap},k} \geq 0, \label{eq:t_lap_min}
\vspace{-5pt}	\end{equation}
\end{small}%
\endgroup    
we can model \emph{driven} laps and \emph{non-driven} laps, thereby jointly optimizing the laps and the number of laps remaining until charging.

To capture the battery energy during the current stint, we keep track of a time to full charge $t_\mathrm{fc}$, defined as
\par\nobreak\vspace{-5pt}
\begingroup
\allowdisplaybreaks
\begin{small}
	\begin{equation}
	t_\mathrm{fc}(j+1) = t_\mathrm{fc}(j) + \Delta t_\mathrm{charge}(j) \quad \forall j\in [1,\dots, n_\mathrm{laps}], \label{eq:t_fc}
\vspace{-5pt}	\end{equation}
\end{small}%
\endgroup 
and bound it through
\par\nobreak\vspace{-5pt}
\begingroup
\allowdisplaybreaks
\begin{small}
	\begin{alignat}{2} 
		&t_\mathrm{fc}(j)\geq 0 \label{eq:t_fc_min}\\
		&t_\mathrm{fc}(j)\leq t_\mathrm{charge,max}, \label{eq:t_fc_max}
	\vspace{-5pt}\end{alignat} 
\end{small}%
\endgroup
where $t_\mathrm{charge,max}$ represents the charge time corresponding to charging the battery from the lower energy bound to the upper energy bound. Since we directly linked this equivalent charge time to the battery energy, we are guaranteed to remain within the battery energy operating limits. To account for the current battery energy, we initialize the time to full charge as
\par\nobreak\vspace{-5pt}
\begingroup
\allowdisplaybreaks
\begin{small}
	\begin{equation}
		t_\mathrm{fc}(1) = t_\mathrm{fc,meas}, \label{t_fc0}
\vspace{-5pt}	\end{equation}
\end{small}%
\endgroup 
where $t_\mathrm{fc,meas}$ is the time to full charge calculated from the current measured battery energy. 

Finally, we ensure that the in-lap is part of the optimal solution and order the driven lap vector with non-driven and driven laps first and last, respectively, with			
\par\nobreak
\begingroup
\allowdisplaybreaks
\begin{small}
	\begin{equation} \label{eq:b_lap}
		b_{\mathrm{lap}}(j+1) \geq b_{\mathrm{lap}}(j),  \quad  \forall j \in [1,n_{\mathrm{laps}}].%for any k >0 
	\end{equation}
\end{small}%
\endgroup

% Show sorting of b?

% Show other remaining constraints?

\subsection{Future Stints}
The remainder of the endurance race is captured on a stint basis, whereby we use stint time maps as a function of the stint length and the charge time, taken from~\cite{KampenHerrmannEtAl2023}. Similar to the current stint, we keep track of the total elapsed time $t_\mathrm{tot}$ through
\par\nobreak\vspace{-5pt}
\begingroup
\allowdisplaybreaks
\begin{small}
	\begin{equation} \label{eq:t_tot}
		t_\mathrm{tot}(i+1) =t_\mathrm{tot}(i) + t_\mathrm{stint}(i) + t_\mathrm{charge}(i) \; \forall i\in [2,\dots,n_\mathrm{stops}-1],
	\end{equation}
\end{small}%
\endgroup   
whereby $t_\mathrm{stint}$ is the time to complete the stint and $t_\mathrm{charge}$ is the charge time after stint. Since we do not have a pit stop after the final stint, the elapsed time after the final stint is defined as
\par\nobreak\vspace{-5pt}
\begingroup
\allowdisplaybreaks
\begin{small}
	\begin{equation} \label{eq:t_elapsed_final}
		t_{\mathrm{tot}}(n_\mathrm{stops}+1) = t_\mathrm{tot}(n_\mathrm{stops}) + t_{\mathrm{stint}}(n_\mathrm{stops}).
	\vspace{-5pt}\end{equation}
\end{small}%
\endgroup 
Because endurance races are time-limited, we bound the total time by
\par\nobreak\vspace{-10pt}
\begingroup
\allowdisplaybreaks
\begin{small}
	\begin{alignat}{2} 
		&t_\mathrm{tot}(j)\geq 0 \label{eq:t_tot_min}\\
		&t_\mathrm{tot}(j)\leq t_\mathrm{race} - t_\mathrm{meas}, \label{t_tot_max}
\vspace{-20pt}	\end{alignat} 
\end{small}%
\endgroup
whereby $t_\mathrm{race}$ represents the race time limit and $t_\mathrm{meas}$ is the current time. We link the current stint and future stints by setting the initial total time to
\par\nobreak\vspace{-5pt}
\begingroup
\allowdisplaybreaks
\begin{small}
	\begin{equation}\label{eq:t_tot0}
		t_\mathrm{tot}(1) =t_\mathrm{stint,current} + t_\mathrm{fc}(n_\mathrm{laps}+1).
\vspace{-5pt}	\end{equation}
\end{small}%
\endgroup 
We model the individual stint times as a positive semi-definite constraint~\cite{BoydVandenberghe2004} as
%\par\nobreak
%\begingroup
%\allowdisplaybreaks
%\begin{small}
%	\begin{equation} \label{eq:t_stint_cone_M}
%		\begin{split}
%			M\cdot (1-b_{\mathrm{pit}}(i)) + t_{\mathrm{stint}}(i) + t_{\mathrm{charge}}(i)  \geq  \\
%			\begin{Vmatrix}
%				2\cdot z_{\mathrm{s}}(i) \\
%				M\cdot (1-b_{\mathrm{pit}}(i)) +t_{\mathrm{stint}}(i) - t_{\mathrm{charge}}(i)   \\	
%			\end{Vmatrix}_2,\\
%		\forall i\in[1,\dots,n_\mathrm{stops}],
%		\end{split}
%	\end{equation}
%\end{small}%
%\endgroup
\par\nobreak
\begingroup
\allowdisplaybreaks
\begin{small}
	\begin{equation}
		t_{\mathrm{stint}}(i) \geq x_{\mathrm{s}}(i)^\top Q_{\mathrm{s}}x_{\mathrm{s}}(i) - M\cdot (1-b_{\mathrm{pit}}(i)),
		\label{eq:t_stint_M}
	\end{equation}
\end{small}%
\endgroup
where $x_{\mathrm{s}}(i) = \left[\frac{1}{\sqrt{t_{\mathrm{charge}}(i)}} \  \sqrt{t_{\mathrm{charge}}(i)} \  \frac{N_{\mathrm{laps}}(i)}{\sqrt{t_{\mathrm{charge}}(i)}}\right]^\top$ and $Q_{\mathrm{s}} \in \mathbb{S}_+^3$ is a symmetric positive semi-definite matrix of coefficients. Since it is optimal to minimize stint time, this constraint will hold with equality at the optimum. For further information on the derivation, we refer the reader to~\cite{KampenHerrmannEtAl2023}. The final stint of the race is not followed by a pit stop in which the battery is charged, which means that the battery can be fully depleted. Therefore, we separately model the final stint with a fixed energy budget corresponding to $t_\mathrm{charge}(n_\mathrm{stops}+1)=t_\mathrm{charge,max}$. With the charge time being fixed, we can then model the final stint time by a quadratic function with the stint length as 
\par\nobreak\vspace{-5pt}
\begingroup
\allowdisplaybreaks
\begin{small}
	\begin{equation}
		t_{\mathrm{stint}}(n_\mathrm{stops}+1) \geq D_\mathrm{s,f}^\top x_\mathrm{s,f},
		\label{eq:t_stint_final}
\vspace{-5pt}	\end{equation}
\end{small}%
\endgroup
where $D_\mathrm{s,f}$ is a vector of coefficients and $x_\mathrm{s,f} = [N_\mathrm{laps}^2(n_\mathrm{stops}+1) \  N_\mathrm{laps}(n_\mathrm{stops}+1) \ 1]^\top$. Whenever a non-driven stint is taken instead of a real stint, i.e., $b_\mathrm{pit}(i) = 0$, we define an upper bound on stint length as
\par\nobreak\vspace{-5pt}
\begingroup
\allowdisplaybreaks
\begin{small}
	\begin{equation} \label{eq:Nlaps_M}
		N_{\mathrm{laps}}(i) \leq N_{\mathrm{laps,max}} \cdot b_{\mathrm{pit}} (i),
\vspace{-5pt}	\end{equation}
\end{small}%
\endgroup
thereby excluding it from the objective function.

Finally, we ensure that the final stint is part of the optimal solution by writing				
\par\nobreak
\begingroup
\allowdisplaybreaks
\begin{small}
	\begin{equation} \label{eq:b_pit}
		b_{\mathrm{pit}}(i+1) \geq b_{\mathrm{pit}}(i),  \quad  \forall i \in [1,n_{\mathrm{stops}}].%for any k >0
	\end{equation}
\end{small}%
\endgroup

% x t_tot as current time over race
% x charge time of first stop from t_fc(n_laps)
% stint map
% charge time per stint
% final stint non integer and different map. Non-integer part already mentioned
% 
%
%\subsection{Last stint} %?
%% charge time current stint if in last stint
%% t_last
		
\subsection{Online Race Strategy Optimization Problem}
				\label{sec:Methodology - MPC optimization}
				This section presents the online race strategy optimization problem of the electric race car. Given a predefined race time, current race time and current battery energy we formulate the control problem using the state variables $x=(t_\mathrm{tot},t_\mathrm{fc})$ and the control variables $u=(\Delta t_\mathrm{charge},b_{\mathrm{pit}},t_{\mathrm{charge}},N_{\mathrm{laps}},b_{\mathrm{pit}})$ as follows:
				\vspace{-10pt}
				\begin{prob}[Maximum-race-distance Strategies]\label{prob:high-level}
					The maximum-race-distance strategies are the solution of
					\par\nobreak\vspace{-5pt}
					\begingroup
					\allowdisplaybreaks
					\begin{small}
						\begin{equation*}
							\begin{aligned}
								& \max \sum_{i=1}^{n_{\mathrm{stops}}} S_{\mathrm{lap}}\cdot N_{\mathrm{laps}}(i) + \sum_{j=1}^{n_{\mathrm{laps}}} S_{\mathrm{lap}}\cdot b_{\mathrm{lap}}(j) ,\\
								&\textnormal{s.t. }  \eqref{eq:t_stint_current}-\eqref{eq:b_pit}. \\
							\end{aligned}
						\vspace{-5pt}\end{equation*}
					\end{small}%
					\endgroup
				\end{prob}
				
				\noindent Problem 1 is a mixed-integer second-order conic program that can be solved with global optimality guarantees~\cite{Lee2012,BelottiKirchesEtAl2013}.
\vspace{-10pt}
\section{Competitor Interactions} \label{sec:competitor interaction}
\begin{figure}
	\vspace{4pt}
	\centering 
	\includegraphics[width=0.7\columnwidth]{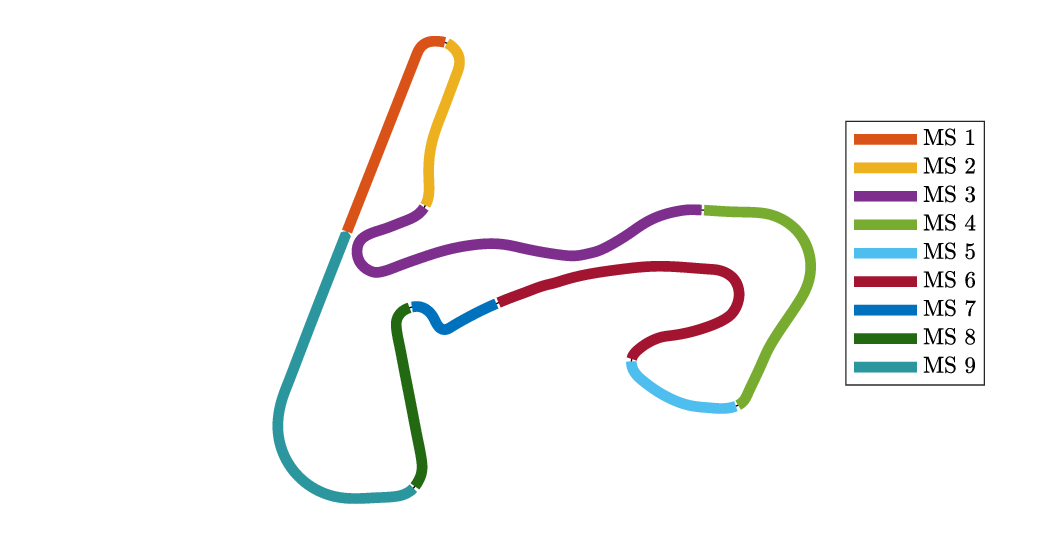}
	\caption{Mini-sector definition for the Zandvoort reference track.}\vspace{-10pt}
	\label{fig:track_MS}
\end{figure}

This section analyzes various types of competitor interactions and model them with time penalties representing possible actions.
During a race, there are many other vehicles driving along the track, each of them with a different pace. These competitors can then act as disturbances on the optimal race strategy whenever our ego vehicle gets in close proximity. In this analysis, we consider three main interactions, being the ego vehicle approaching a competitor (\textit{Attack}), the ego vehicle being approached by a competitor from behind (\textit{Defend}) and track position prediction after a pit stop (\textit{Pit-lane Exit Traffic}).
For each of these interactions, there is a set of actions that can be taken by the ego vehicle. The optimal decision then consists of the action with the lowest corresponding estimated time penalty.
The time penalties are estimated with a limit of one single interaction considered per \gls{acr:MS} $m$.
We decompose the track into a total of 9 \gls{acr:MS} based on the three main sectors and the most common locations with interactions from previous races. As an example, Fig.~\ref{fig:track_MS} shows the \gls{acr:MS} for the Zandvoort track.

\subsection{Attack}
\label{ss:attack}
The \textit{Attack} interaction happens when the ego vehicle has the chance to overtake the car in front during the upcoming \gls{acr:MS}. This is triggered by predicting the virtual gap at the end of it, e.g., if the ego vehicle is predicted to be in front at the end of the \gls{acr:MS} according to the free-flow strategy, we are \textit{virtually overtaking} the car ahead and the \textit{Attack} decision-making process is triggered. The set of possible actions by the ego vehicle is composed by \textit{Overtake}, \textit{Stay Behind} and \textit{Box}, while the competitor's decision is always assumed to be the most aggressive one, i.e., \textit{Defend}.

%
% This is shown in Fig.~\ref{fig:attack_trigger}, where $ms$ indicates the current checkpoint and  $\Delta T_\mathrm{p}$ the predicted gap, which is positive when the competitor is behind and negative when ahead. 
%\begin{figure}[]
%	\centering
%	\footnotesize
%	\input{Figures/attack_trigger}
%	\caption{Virtual overtake.}
%	\label{fig:attack_trigger}
%\end{figure}

\subsubsection{Overtake}\label{sec:overtake}
We define the time penalty $t_\mathrm{pen,ov}$ as the sum of fixed time losses from the \textit{Overtake} attempt $t_\mathrm{ov,c}$, time gain upon a successful overtake $t_\mathrm{ov,g}$ and time losses upon a failed overtake $t_\mathrm{ov,l}$:
%\par\nobreak
%\begingroup
%\allowdisplaybreaks
%\begin{small}
%	\begin{alignat}{2} 
%		\label{eq:T_pen_ov}
%t_\mathrm{pen,ov}(\Delta E_\mathrm{b},s)&= \frac{(t_\mathrm{ov,c}(\Delta E_\mathrm{b})-t_\mathrm{ov,g}(\Delta E_\mathrm{b},s)+ t_\mathrm{ov,l}(s) )}{P_\mathrm{ov}(\Delta t_\mathrm{d},s)}, \\
%t_\mathrm{ov,c}(\Delta E_\mathrm{b}) &= (\Delta t_\mathrm{charge,ov}(\Delta E_\mathrm{b})+t_\mathrm{rl},\\
%t_\mathrm{ov,g}(\Delta E_\mathrm{b},s) &= \left(\Delta t_\mathrm{d}(\Delta E_\mathrm{b})-\Delta t_\mathrm{p}(s)\right)\cdot P_\mathrm{ov}(\Delta t_\mathrm{d},s),\\
%t_\mathrm{ov,l}(s) &=\left(\Delta t_\mathrm{p}(s)+t_\mathrm{gap,min}\right)\cdot \left(1-P_\mathrm{ov}(\Delta t_\mathrm{d},s)\right), \label{eq:T_pen_ov_end}
%	\end{alignat} 
%\end{small}%
%\endgroup
\par\nobreak
\begingroup
\allowdisplaybreaks
\begin{small}
	\begin{equation}
	t_\mathrm{pen,ov}(\Delta E_\mathrm{b},m)= \frac{(t_\mathrm{ov,c}(\Delta E_\mathrm{b})-t_\mathrm{ov,g}(\Delta E_\mathrm{b},m)+ t_\mathrm{ov,l}(m) )}{P_\mathrm{ov}(\Delta t_\mathrm{d},m)},
	\end{equation}
	\begin{alignat}{2} 
		\label{eq:T_pen_ov}
%		&t_\mathrm{pen,ov}(\Delta E_\mathrm{b},m)= \frac{(t_\mathrm{ov,c}(\Delta E_\mathrm{b})-t_\mathrm{ov,g}(\Delta E_\mathrm{b},m)+ t_\mathrm{ov,l}(s) )}{P_\mathrm{ov}(\Delta t_\mathrm{d},m)}, \\
			t_\mathrm{ov,c}(\Delta E_\mathrm{b}) &= \Delta t_\mathrm{charge,ov}(\Delta E_\mathrm{b})+t_\mathrm{rl},\\
			t_\mathrm{ov,g}(\Delta E_\mathrm{b},m) &= \left(\Delta t_\mathrm{d}(\Delta E_\mathrm{b})-\Delta t_\mathrm{p}(m)\right)\cdot P_\mathrm{ov}(\Delta t_\mathrm{d},m),\\
			t_\mathrm{ov,l}(m) &=\left(\Delta t_\mathrm{p}(m)+t_\mathrm{gap,min}\right)\cdot \left(1-P_\mathrm{ov}(\Delta t_\mathrm{d},m)\right), \label{eq:T_pen_ov_end}
	\end{alignat} 
\end{small}%
\endgroup
where $\Delta t_\mathrm{charge,ov}$ captures the increase in charge time due to the additional battery energy used, $t_\mathrm{rl}$ is the time penalty for deviating from the optimal racing line, $\Delta t_\mathrm{d}$ and $\Delta t_\mathrm{p}$ are the desired and predicted gaps with the competitor, respectively, $P_\mathrm{ov}$ is the probability of a successful overtake and $t_\mathrm{gap,min}$ is the minimum following gap to a competitor in front. For the time gain and loss estimation, we essentially take the gap w.r.t. the original strategy and weigh it by the respective probability. Finally, we divide by $P_\mathrm{ov}$ to prevent overtakes whenever the chance of succeeding is minor. The probability of a successful overtake both depends on the predicted gap to the competitor and the track layout, and is estimated from previous race results. Since we can choose to drive faster, thereby increasing the odds of a successful overtake, we jointly optimize the time penalty and the vehicle operation using a modified minimum-lap-time framework~\cite{KampenHerrmannEtAl2023} and adding~\eqref{eq:T_pen_ov}-\eqref{eq:T_pen_ov_end} as constraints to obtain a perturbation around the original solution.

\subsubsection{Stay Behind}
The aforementioned \textit{Overtake} maneuver potentially requires more energy, which means that staying behind the competitor and saving energy can be beneficial. The time penalty $t_\mathrm{pen, sb}$ is computed as

% Complete equation with correct symbols etc. 

% Danger is to stay behind for too many mini sectors and lose a lot of time, but only to find out at the end of the lap when the MPC is triggered again. Therefore, we use a buffer/cumulative time lost

% T_buff: increases when staying behind is decided. Also when Tp<0, we predict to be behind the competitor. However, it can happen that within a MS we still catch the opponent, without overtaking. Then we still lose time and this is added to T_buff. t_buff is reset whenever the interaction ends: overtake or box from either side.
% The time lost in the next MS is also added to provide some lookahead, otherwise we could decide to stay behind in the current MS and lose little time, but lose a lot in the next one because we cannot overtake there.
\par\nobreak
\begingroup
\allowdisplaybreaks
\begin{small}
	\begin{equation}
		\begin{split}
		\label{eq:T_pen_sb}
		t_\mathrm{pen, sb}(\Delta E_\mathrm{b},m)=\Delta t_\mathrm{charge,sb}(\Delta E_\mathrm{b})+\Delta t_\mathrm{p}(m)+t_\mathrm{gap,min}\\+t_\mathrm{buff,sb}(m)+\Delta t_\mathrm{p}(m+1),
	\end{split}
	\end{equation} 
\end{small}%
\vspace{-10pt}
\begin{small}
	\begin{align}
		t_\mathrm{buff,sb}(s+1)=\begin{cases}
			t_\mathrm{buff,sb}(s) + t_\mathrm{pen,sb}(\Delta E_\mathrm{b},m), & \text{if stay behind}\\
			0,		  									   & \text{otherwise}
		\end{cases}
	\end{align} 
\end{small}%
\endgroup
%\begin{equation}
%	\begin{split}
%		\label{eq:T_pen_sb}
%		t_\mathrm{pen, sb}=&\left(\Delta t_\mathrm{charge,sb}+\Delta t_\mathrm{p}+t_\mathrm{gap,min}\right)\\&+t_\mathrm{buff,sb}+t_\mathrm{loss,next\ MS},
%	\end{split}
%\end{equation}
where $\Delta t_\mathrm{charge,sb}$ is the decrease in charge time due to energy saved and $t_\mathrm{buff,sb}$ is a time buffer that keeps track of the total time lost during the \textit{Stay Behind} action. Since the race strategy is only updated every lap, it could occur that a considerable amount of time is lost if the \textit{Stay Behind} action is chosen for consecutive \gls{acr:MS}. Yet this would be unnoticed until the race strategy is triggered. We prevent this by adding an accumulating buffer, capturing the total time lost during the action, instead of only the local time loss. Furthermore, we include the predicted gap of the consecutive \gls{acr:MS} to add more incentive to overtake in the case where this consecutive \gls{acr:MS} provides little overtaking opportunities. We again optimize the vehicle operation to obtain the minimum time penalty for this action.  

%This last term is computed only if  $P_\mathrm{ov,ms}$ of the next MS is lower than the current one: in this way, we consider that not every part of the track allows an overtake. Once again, a modified version of the minimum-energy-lap framework is employed to compute the energy saved by slowing down, also considering the slipstream of the leading car, which is implemented as a constraint by making drag and lift coefficients as variables; finally, initial and final speeds are imposed, as previously explained. In the last stint, $\Delta T_\mathrm{E,sb}$ is used instead $\Delta T_\mathrm{charge,sb}$: this is computed in the same way as in \textit{Overtake}, but more energy is available for the remaining laps.\\
%Even if the objective is convex, the constraints do not define a convex set. Thus, the problem can be solved with a non-linear solver, where the speed is enhanced through a warm-start.

\subsubsection{Box}\label{sec:Box}
Whenever there is an interaction in the final \gls{acr:MS}, we can decide to prematurely stop to charge the vehicle, thereby avoiding possible time loss from the interaction. However, since this often means deviating from the race strategy, we have to evaluate the impact on the race by solving Problem~\ref{prob:high-level} and enforcing the current lap to be an in-lap. If the distance covered exceeds the distance covered in the original solution, this action can be beneficial. In motorsports, this is often referred to as \emph{undercut} and is most likely to be viable when vehicles have a similar pace and when we are close to making a pit stop.
% in essence we rerun the MPC, forcing a pit stop at the end of the current lap. If it results in a larger distance covered, we take it.
% This action is often referred to as the undercut and can be viable when the ego vehicle and the competitors have very similar pace (for example in single car classes), since it allows for free flow laps after the pit stop.
%
%
%
%If the interaction happens during the last MS, it is possible to analyze the convenience of an anticipated pit-stop. Obviously, by changing the pit-stop lap, we are making the race strategy sub-optimal: it is necessary to quantify the impact to evaluate this action. First, the last MS is re-optimized, imposing to enter the pit-lane; then, the race strategy is recomputed through the MPC, starting from an OUT-lap. To evaluate the penalty related to this decision, we first compare the number of laps: if this is lower for the new computed strategy, the penalty is set to a high value; otherwise, this is computed as
%\begin{equation}
%	\label{eq:T_pen_box}
%	T_\mathrm{penalty, box}=T'-T_\mathrm{box}',
%\end{equation}
%where $T_\mathrm{box}'$ is computed for the modified strategy, while $T'$ is related to the original one. This term is graphically explained in Fig.~\ref{fig:T'}.
%\begin{figure}[]
%	\centering
%	\footnotesize
%	\input{Figures/T_box_expl}
%	\caption{Graphical explanation of $T'$.}
%	\label{fig:T'}
%\end{figure}
%This decision is analyzed only during the last MS of laps closed to the optimal pit-stop lap. The time penalty is set to a large value during the last stint since no more stops are planned.

\subsection{Defend}
Similar to the \textit{Attack} interaction, the \textit{Defend} interaction occurs when we get \textit{virtually overtaken} by a competitor from behind, i.e., the predicted gap between the ego vehicle and the competitor $t_\mathrm{p}$ is negative at the end of the next \gls{acr:MS}. The set of possible actions is composed by \textit{Block}, \textit{Let Through} and \textit{Box}, where the latter is the same as explained in Section~\ref{sec:Box}. We assume that the competitor is always taking the most aggressive action, i.e., \textit{Overtake}.
%	\begin{figure}[]
	%		\centering
	%		\footnotesize
	%		\input{defence_trigger}
	%		\caption{Virtual overtake - Defence.}
	%		\label{fig:def_virt_ov}
	%	\end{figure}

\subsubsection{Block}
Similar to the \textit{Overtake} action, the \textit{Block} action impact $t_\mathrm{pen,blk}$ is defined as a sum of fixed time losses $t_\mathrm{blk,c}$, time gain upon a successful blocking maneuver $t_\mathrm{blk,g}$ and time losses upon a failed attempt $t_\mathrm{blk,l}$:
\par\nobreak\vspace{-5pt}
\begingroup
\allowdisplaybreaks
\begin{small}
	\begin{equation}
		t_\mathrm{pen,blk}(\Delta E_\mathrm{b},m)= \frac{(t_\mathrm{blk,c}(\Delta E_\mathrm{b})-t_\mathrm{blk,g}(\Delta E_\mathrm{b},m)+ t_\mathrm{blk,l}(m) )}{P_\mathrm{def}(\Delta t_\mathrm{d},m)}, 
	\end{equation}
\begin{fleqn}
	\begin{alignat}{1} 
		\label{eq:T_pen_def}
%		t_\mathrm{pen,blk}(\Delta E_\mathrm{b},m)&= \frac{(t_\mathrm{blk,c}(\Delta E_\mathrm{b})-t_\mathrm{blk,g}(\Delta E_\mathrm{b},m)+ t_\mathrm{blk,l}(s) )}{P_\mathrm{def}(\Delta t_\mathrm{d},m)}, \\
		&t_\mathrm{blk,c}(\Delta E_\mathrm{b}) = (\Delta t_\mathrm{charge,blk}(\Delta E_\mathrm{b})+t_\mathrm{rl},\\
		&t_\mathrm{blk,g}(\Delta E_\mathrm{b},m) = \left(\Delta t_\mathrm{d}(\Delta E_\mathrm{b})\right)\cdot P_\mathrm{def}(\Delta t_\mathrm{d},m),\\
		&t_\mathrm{blk,l}(m) =\max \left(\Delta t_\mathrm{p}(m)+t_\mathrm{gap,min},0\right)\cdot \left(1-P_\mathrm{def}(\Delta t_\mathrm{d},m)\right), \label{eq:T_pen_blk_end}
\vspace{-5pt}	\end{alignat} 
\end{fleqn}
\end{small}%
\endgroup
%\begin{equation}
%	\label{eq:T_pen_def}
%	\begin{split}
%		t&_\mathrm{pen,blk}(\Delta E_\mathrm{b})=\\&(\Delta t_\mathrm{charge,blk}(\Delta E_\mathrm{b})+t_\mathrm{rl}-\Delta t_\mathrm{d,blk}\cdot P_\mathrm{def}\\
%		&+\max\left(t_\mathrm{gap,min}+\Delta t_\mathrm{p},\ 0\right)\cdot (1-P_\mathrm{def}))\cdot \frac{1}{P_\mathrm{def}},
%\end{split}\end{equation}
%\begin{figure}[]
%	\centering
%	\footnotesize
%	\input{Figures/dT_trg_def}
%	\caption{Graphical explanation of $\Delta T_\mathrm{trg,def}$.}
%	\label{fig:dT_trg_def}
%\end{figure}
% Instead of using P_ov_c, we define P_def as the probability of a successful defend and not the probability of a sucessful overtake by the competitor
% t_d is with respect to the original strategy
where $\Delta t_\mathrm{charge,blk}$ is the additional charge time due to the defending maneuver and $P_\mathrm{def}$ is the probability of successfully blocking the competitor, thereby preventing the ego vehicle from being overtaken. Note that the definition of this time penalty is similar to the penalty defined in Section~\ref{sec:overtake}, with the addition of a saturation for the time losses as $\Delta t_\mathrm{p}\leq0$ during \textit{Defend} interactions. Again, to find the optimal trade-off between driving faster and increasing the probability of a successful block, and the energy consumption, we optimize the vehicle operation. 

%competitor's probability to overtake our car, defined as $P_\mathrm{ov,c}=P_\mathrm{ov,c,\ comp}\cdot P_\mathrm{ov,\ ms}$, similarly to what already explained for the overtaking maneuver. The term $\max\left(T_\mathrm{beh}+\Delta T_\mathrm{p},\ 0\right)$ quantifies the time lost by slowing down to respect the minimum time distance at which another car can be followed.\\
%To minimize the probability of losing the position, it is necessary to minimize $P_\mathrm{ov,c}$. Being this term dependent on $\Delta T_\mathrm{trg,def}$, this is obtained by improving the MS time, i.e. increasing $\Delta T_\mathrm{trg,def}$. On the other hand, by pushing more, we also consume more energy, increasing  $\Delta T_\mathrm{charge,def}$.\\
%To find the optimum trade-off, we develop an optimization problem through an adapted minimum-energy-lap framework, where constraints related to the probability of being overtaken, energy consumption and initial and final speeds have been added.	During the last stint, $\Delta T_\mathrm{charge,def}$ is replaced by $\Delta T_\mathrm{E,def}$, as previously explained for the other time penalties.\\
%The problem can be solved with a non-linear solver, being the objective function non-convex. A warm start is employed to enhance the computation speed.

\subsubsection{Let Through}
Instead of attempting to block the competitor, the ego vehicle can also decide to continue following the original strategy and let the competitor pass. Although this action might seem counter-intuitive in the first place, this can be very beneficial on the long term. The blocking action discussed in the previous section can require a considerable amount of additional energy if the competitor is significantly faster. Instead, it can be beneficial to follow the original strategy and let the competitor pass, which is also often seen in races involving pit stops. For example, if the ego vehicle spends more energy on preventing the \textit{Overtake}, it can occur that it has to charge one lap earlier and thereby losing the interaction. In this case, the energy is essentially wasted, since the position is not preserved. To detect interference with the original strategy of the ego vehicle after a pass from the competitor, we check for possible interactions in the three upcoming \gls{acr:MS} after. If no other interaction occurs with the same competitor, we always decide to let the competitor pass. However, if another interaction is predicted, we have to estimate the time penalty of letting the competitor through and staying behind for the entire horizon.  
% What is often seen in motorsport classes that involve pit stops is that instead of trying to prevent the competitor from overtaking the ego vehicle, the competitor is let through, thereby saving energy and minimizing the deviation from the original optimal strategy.
% Therefore we provide some lookahead for the 3 upcoming MS and check if a possible interaction could happen if the competitor is let through. If we do not reach the competitor, it is always beneficial to let through, since this allows us to stay closest to the original strategy. However, if another interaction is predicted in the horizon, we have to calculate the time penalty to see whether it is better to defend now or to let through and overtake later, including this horizon.

% Check in upcoming MS if we interact again with competitor.
% If we do, it means that with original strategy we catch it again. This means that we estimate the time penalty by deciding to stay behind, since reovertaking does not make sense again after a let through decision within the upcoming 3 MS. Then we  set target time such that we let through and stay exactly behind. We can do this, since we are predicted to catch anyway, so following should always be possible.
% Staying behind is applied by default since it does not make sense to
% overtake opponent again after letting it pass.

The time penalty is defined as the difference between the original strategy $t_{\mathrm{MS, p}}$ and the target time $t_{\mathrm{target}}$, where the latter is set such that we follow the competitor:
\par\nobreak
\begingroup
\allowdisplaybreaks
\begin{small}
	\begin{equation}
		\begin{split}
		\label{eq:T_pen_lt}
t_\mathrm{pen,lt}(\Delta E_\mathrm{b},m)=&\max\left(t_\mathrm{gap,min}+\Delta t_\mathrm{p}(m),0\right)+\\ \sum_{l=s+1}^{s+3}(t_{\mathrm{target}}(l)&-t_{\mathrm{MS, p}}(l) +\Delta t_{\mathrm{charge,lt}}(\Delta E_\mathrm{b},l)),
\end{split}
	\end{equation} 
\end{small}%
\endgroup
%\begin{equation}
%	\begin{split}
%		\label{eq:T_pen_lt}
%		T&_\mathrm{pen,lt}=\max\left(T_\mathrm{beh}+\Delta T_\mathrm{p},\ 0\right)\\&+\sum_{m=MS+1}^{m+2}(T_{\mathrm{target}}(m)-T_{\mathrm{MS, act}}(m)\\
%		& +\Delta T_{\mathrm{charge,lt}}(m)),
%	\end{split}
%\end{equation}	
where $\Delta t_{\mathrm{charge,lt}}(\Delta E_\mathrm{b},m)$ is the change in charging time w.r.t. the original strategy in \gls{acr:MS} $m$. The change in battery energy is then computed by optimizing the vehicle operation such that the target \gls{acr:MS} are met. % and $t_{\mathrm{target}}(m)-t_{\mathrm{MS, p}}(m)$ is the pace difference compared to the original predicted \gls{acr:MS} time. 
% 
%The time penalty is computed by optimizing the three MS after the upcoming one. Particularly, we slow down where we are faster and keep the target pace where we are slower. Once again, the optimization problem is a modified minimum-energy-lap framework, where the initial and final speeds of the first and the last considered MS and the aerodynamic constraints are imposed. Also in this case, during the last stint, we use  $\Delta T_\mathrm{E,lt}$ instead of $\Delta T_{\mathrm{charge,lt}}$, as previously explained.\\
%The problem is solved using a non-linear solver since the constraints do not define a convex set. Once again, the computation speed is enhanced by a warm start.

\subsection{Pit-exit Traffic} 
% Possibly important
% What can we do to avoid? -> shorten charging or overtake during the out lap.
% We could of cours overtake the competitor during the out-lap. However, another action we can consider is to stop earlier with charging in order to exit the pit lane in front of the competitor. 

After a charging stop and exiting the pit-lane, it can occur that the ego vehicle returns to the track behind a slower competitor, resulting in a possible time loss. This scenario is referred to as pit-exit traffic and it is a common challenge in motorsports. Of course, it is possible to attempt to \textit{Overtake} the competitor during the out-lap, but it could be advantageous to shorten the charging stop and attempt to arrive in front of the competitor. Hereby, we would trade charged energy for a better track position. To analyze the interaction, we compare the decision of \textit{Shorten Pit-stop} with \textit{Overtake during Out-lap}, assuming that the competitor is always defending the position.

%The pit-exit traffic is a critical scenario in motorsport. In fact, exiting the pit-lane behind a slower car could result in a significant time loss: this can be avoided by having a shorter pit-stop, which means charging less the battery. Hence, the possible set of decisions is composed of \textit{Short Pit-stop} and \textit{Overtake during OUT-lap}, while we assume that the competitor is always defending.

\subsubsection{Short Pit-stop}
% Continue here
% If it is not the last stint, we just take the Tp with margin as time penalty, since this is the time that we would have to be charging longer in the next pit stop.
%  However, this is actually wrong since the lower level energy limit stays the same, so in reality we just have less energy abvailable.
% So what we should actuelly do is to claculate the lap times with this reduced amount of energy, as is done for the last stint.
% 
To investigate whether a shorter pit stop is beneficial, we have to check the increase in lap time due to the reduced amount of battery energy available. Therefore, the time penalty $t_\mathrm{pen,sp}$ is defined as the difference in lap time between the original strategy and the predicted lap times with the reduced battery energy, whereby we assume to spread the gap in battery energy across the entire stint:
\par\nobreak
\begingroup
\allowdisplaybreaks
\begin{small}
	\begin{equation}
			\label{eq:T_pen_sp}
			t_\mathrm{pen,sp}=\sum_{j=1}^{N_\mathrm{laps}}\left(t_\mathrm{lap}\left(\Delta \overline{t}_\mathrm{charge}(j)- \frac{\Delta t_\mathrm{charge,sp}}{N_\mathrm{laps}} \right)-\overline{t}_\mathrm{lap} \right),
	\end{equation} 
\end{small}%
\endgroup
where $\Delta \overline{t}_\mathrm{charge}$ is the equivalent charge time per lap in the original strategy, $\overline{t}_\mathrm{lap}$ is the predicted lap time of the original strategy, $\Delta t_\mathrm{charge,sp}$ is the deduction of charge time needed in order to exit the pit-lane in front of the competitor and $N_\mathrm{laps}$ is the predicted stint length. In this case, we do not need an optimization framework, since we can use pre-computed lap time maps.

\subsubsection{Overtake during the Out-lap}
In the case where the ego vehicle is predicted to catch up with the competitor during the out-lap, we calculate the time penalty for an \textit{Overtake} maneuver as described in \mbox{Section~\ref{ss:attack}.}

%
%
%In this section, we present the online race strategy optimization problem. First, we formulate the maximum-race-distance control problem that optimizes the stint length and charge time for a pre-defined number of pit stops. Second, we model the minimum stint time by leveraging the low-level control problem and optimizing for various combinations of stint length and initial battery energy, as was shown in Fig.~\ref{fig:framework}.
%Finally, we extend the maximum-race-distance control problem to allow joint optimization of the stint length, charge time, and number of pit stops. 

\section{Results } \label{Results}

This section presents the numerical results for the combined MPC framework and competitor interaction decision making. As a use case, we choose the InMotion electric endurance race car as our ego vehicle~\cite{InMotion} and the 2023 Supercar Challenge at the Zandvoort circuit as the race event. This event consists of a \unit[1]{h} race with a total of \unit[31]{participants}, where all other cars are equipped with an \gls{acr:ice}.
To simulate the race, we assume that the position and lap times of the competitors are not influenced by the ego vehicle. Furthermore, a decision to overtake by the ego vehicle does not guarantee a successful overtake. Since the interactions are stochastic due to the overtake and defend probabilities, the outcome of the race would be stochastic as well. However, since we cannot analyze the outcome for every possible scenario, we impose the outcome of an interaction to be the outcome with the highest probability. This allows for a fair comparison against a baseline decision making strategy, whereby the ego vehicle will always attempt to overtake a competitor.

The online race strategy problem is parsed using YALMIP~\cite{Loefberg2004} and solved using Gurobi~\cite{GurobiOptimization2021}, while the time penalties and vehicle operation in the \gls{acr:MS} is solved and parsed using CasADi~\cite{Andersson2019} and IPOPT~\cite{IPOPT}, respectively. The average fraction of solver time over the \gls{acr:MS} time is \unit[15]{\%}, while the worst case fraction is \unit[27]{\%} of the corresponding \gls{acr:MS} time on an Intel Core i7-4710MQ 2.5 GHz processor with 8GB of RAM, demonstrating the real-time capabilities of the framework. 

% In this figure we observe gains in position while no overtake action is taken. These overtakes are from competitors pitting or having an accident (since they lose multiple positions).
\begin{figure}
	\vspace{4pt}
	\centering 
	\input{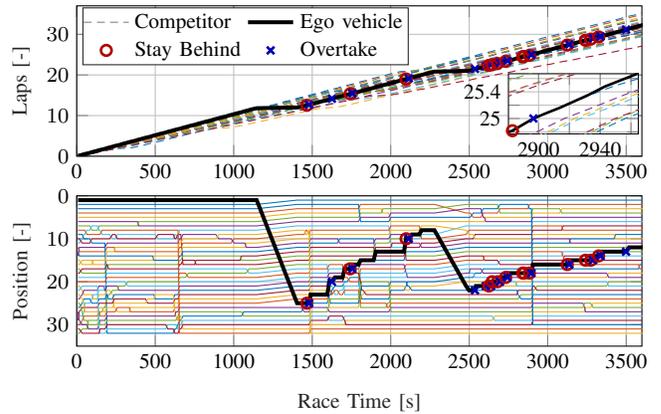}\vspace{-15pt}
	\caption{Evolution of driven distance and position as a function of time for the ego vehicle and competitors, together with a zoom of an overtake. The ego vehicle starts at the first position, but has to charge before the competitors make their pit stop, resulting in a loss of several positions. Thereafter, the ego vehicle starts overtaking the competitors, resulting in the \nth{13} overall finishing position among 32 participants.}\vspace{-10pt}
	\label{fig:race_results}
\end{figure}

The evolution of the race as a function of time is shown in Fig.~\ref{fig:race_results} for both the ego vehicle and the competitors. Since the ego vehicle is capable of performing the fastest lap times, it starts in \nth{1} place. Furthermore, no defensive interactions are observed, since the ego vehicle is the fastest during a lap. At the beginning of the race, the ego vehicle creates as gap to the competitors, until it requires charging. Since the charging time is considerably longer than the pit stop time of the \gls{acr:ice}-driven competitors, the ego vehicle drops back in position, resulting in several interactions for the remainder of the race. Ultimately, the ego vehicle finishes the race in \nth{13} place out of the \unit[32]{participants}, revealing that e-race cars could compete with \gls{acr:ice} race cars in the near future.
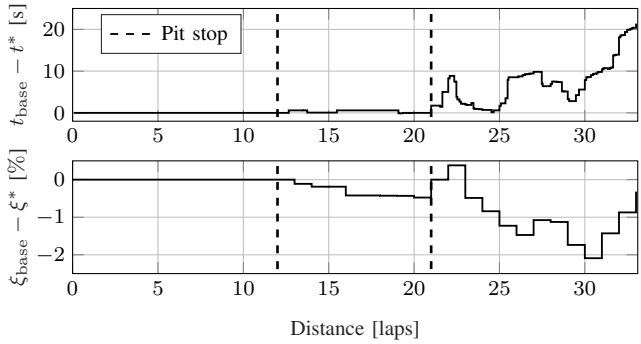
\begin{figure}
	\vspace{4pt}
	\centering 
\definecolor{mycolor1}{rgb}{0.00000,0.44700,0.74100}%
\begin{tikzpicture}
		\footnotesize
		\begin{groupplot}[%
			group style={
				group size=1 by 2,
				vertical sep=15pt},
			width=0.87\columnwidth,
			at={(0,0)},
			scale only axis,
			xmin=0,
			xmax=33.1,
			y label style={yshift=-0.5cm},
			axis background/.style={fill=white},
			xmajorgrids,
			ymajorgrids,
			]
			
			\nextgroupplot[%
			height=1.5cm,
			ymin=-2,
			ymax=25,
			ylabel style={font=\color{white!15!black}},
			ylabel={$t_\mathrm{base}-t^\ast$ [s]},
			axis background/.style={fill=white},
%			axis x line*=bottom,
%			axis y line*=left,
			legend style={at={(0.05,0.95)}, anchor=north west, legend cell align=left, align=left, draw=white!15!black},
			%		legend image post style={line width=0.7pt},
%			legend columns=2,
			xmajorgrids,
			ymajorgrids,
			]
	
	\addplot[const plot, color=black, line width=0.7pt, forget plot] table[row sep=crcr] {%
		0.084526884245129	0\\
		12.6574313219066	0.599999999999909\\
		13.0845268842451	0.599999800262822\\
		13.159661892463	0.59999974819425\\
		13.3521953510214	0.599999357033312\\
		13.4780464897863	0.599999126223338\\
		13.5322845738436	0.599999080433236\\
		13.6574313219066	0.599999019822917\\
		13.7419582061517	0.0649986303621972\\
		13.8170932143696	0.0649985839384044\\
		14	0.0649981024666886\\
		14.3521953510214	0.0649981019621464\\
		15.0845268842451	0.0649979022250591\\
		15.159661892463	0.0649978501564874\\
		15.3521953510214	0.0649974589955491\\
		15.4780464897863	0.599999999999909\\
		15.5322845738436	0.599999954243231\\
		15.6574313219066	0.599999893632912\\
		15.7419582061517	0.599999770752675\\
		15.8170932143696	0.599999724328882\\
		16	0.599999242857166\\
		17.0845268842451	0.599999442594253\\
		17.159661892463	0.599999494662825\\
		17.3521953510214	0.599999885823763\\
		17.4780464897863	0.600000116633737\\
		17.5322845738436	0.600000162423839\\
		17.6574313219066	0.600000223034158\\
		17.7419582061517	0.600000345914395\\
		17.8170932143696	0.600000392338188\\
		18	0.600000873809904\\
		18.0845268842451	0.600000674072817\\
		18.159661892463	0.600000622004245\\
		18.3521953510214	0.600000230843307\\
		18.4780464897863	0.600000000033333\\
		18.5322845738436	0.599999954243231\\
		18.6574313219066	0.599999893633139\\
		18.7419582061517	0.599999770753129\\
		18.8170932143696	0.599999724329336\\
		19	0.599999242857393\\
		19.0845268842451	-0.0998635537202972\\
		19.159661892463	-0.0998506906903458\\
		19.3521953510214	-1.67938196682371e-09\\
		19.4780464897863	5.70176835026359e-05\\
		19.5322845738436	6.83296625538787e-05\\
		19.6574313219066	8.3302818438824e-05\\
		19.7419582061517	0.000113659200906113\\
		19.8170932143696	0.000125127751744003\\
		20	0.000244070656208351\\
		21	1.73731418719262\\
		21.5322845738436	1.44348193298856\\
		21.6574313219066	5.01545957525468\\
		22	8.26127832448901\\
		22.0845268842451	8.86354299787945\\
		22.159661892463	8.86409020015435\\
		22.3521953510214	7.49444009496483\\
		22.4780464897863	3.82553233898989\\
		22.5322845738436	3.22608486661238\\
		22.6574313219066	2.69299999999976\\
		22.7419582061517	2.18884851246185\\
		22.8170932143696	2.18938484817272\\
		23	1.98999999999933\\
		23.0845268842451	1.9178329029005\\
		23.159661892463	1.90060079480236\\
		23.3521953510214	2.36248638758661\\
		23.4780464897863	1.08223460493491\\
		23.5322845738436	0.959234723254212\\
		23.6574313219066	0.938175414214584\\
		23.7419582061517	0.894414035384216\\
		23.8170932143696	0.877931492517746\\
		24	0.704287003602985\\
		24.5322845738436	0.129290560768823\\
		24.6574313219066	0.599999999999909\\
		25	1.69558707413262\\
		25.0845268842451	2.27931081562429\\
		25.3521953510214	3.48914699790339\\
		25.4780464897863	7.92765891008639\\
		25.5322845738436	8.52765891450417\\
		26.0845268842451	8.69479608152506\\
		26.159661892463	8.73156436833642\\
		26.3521953510214	9.03298585877565\\
		26.4780464897863	9.21551015743216\\
		26.5322845738436	9.24817005727891\\
		26.6574313219066	9.2846150271107\\
		26.7419582061517	9.38674661681625\\
		26.8170932143696	9.42178983118583\\
		27	9.83003397389803\\
		27.0845268842451	9.83003397937864\\
		27.159661892463	9.83003398070286\\
		27.3521953510214	9.83003399133804\\
		27.4780464897863	6.99087030298824\\
		27.5322845738436	6.39087030367409\\
		27.6574313219066	6.3908703053844\\
		27.7419582061517	6.39087030889732\\
		27.8170932143696	6.39087031019517\\
		28	6.9426112909091\\
		28.0845268842451	7.49959892253491\\
		28.159661892463	7.4900739353775\\
		28.3521953510214	7.41483985959667\\
		28.4780464897863	7.37003755741671\\
		28.5322845738436	7.36198737338873\\
		28.6574313219066	5.15659140733987\\
		28.7419582061517	5.12905333289291\\
		28.8170932143696	5.12008639927217\\
		29	3.26748832537714\\
		29.0845268842451	2.80716562550879\\
		29.4780464897863	4.25145189466139\\
		29.5322845738436	4.25145189516843\\
		29.6574313219066	5.58114647261209\\
		30	7.84624955457457\\
		30.0845268842451	8.50139357529997\\
		30.159661892463	8.56100986304455\\
		30.3521953510214	9.03875987654965\\
		30.4780464897863	9.31951309161832\\
		30.5322845738436	9.37067965543383\\
		30.6574313219066	9.42802532627729\\
		30.7419582061517	9.74330685587893\\
		30.8170932143696	9.79997300600007\\
		31	10.4200417029429\\
		31.0845268842451	10.6532842244219\\
		31.159661892463	10.7062897081846\\
		31.3521953510214	10.5339432162327\\
		31.4780464897863	10.7918067908799\\
		31.5322845738436	10.8391027727248\\
		31.6574313219066	13.737585640954\\
		31.7419582061517	13.8871195678289\\
		31.8170932143696	13.9374806135643\\
		32	18.0819740830957\\
		32.0845268842451	18.525146026228\\
		32.159661892463	18.6253778626842\\
		32.3521953510214	19.3725301153049\\
		32.4780464897863	19.8315905403288\\
		32.5322845738436	19.9171075914023\\
		32.6574313219066	20.041110210027\\
		32.7419582061517	20.2918321042389\\
		32.8170932143696	20.3879423558606\\
		33	21.3774606755592\\
	};
	
	\addplot [color=black, dashed, forget plot,line width=1pt]
	table[row sep=crcr]{%
		12 -1\\
		12	30\\
	};
\addplot [color=black, dashed, forget plot,line width=1pt]
table[row sep=crcr]{%
	21	-1\\
	21	30\\
};
\addlegendimage{dashed,black,line width=0.7pt}
\addlegendentry{Pit stop}

	\nextgroupplot[%
	height=1.5cm,
	xmin=0,
	xmax=33.1,
	xlabel style={font=\color{white!15!black}},
	xlabel={Distance [laps]},
	ymin=-2.5,
	ymax=0.5,
	ylabel style={font=\color{white!15!black}},
	ylabel={$\xi_\mathrm{base}-\xi^\ast$ [$\%$]},
	axis background/.style={fill=white},
%	axis x line*=bottom,
%	axis y line*=left,
	]
	
	\addplot[const plot, color=black, line width=0.7pt, forget plot] table[row sep=crcr] {%
		0	0\\
		13	-0.112050261102915\\
		14	-0.187727326667044\\
		15	-0.187784640887358\\
		16	-0.421302510324907\\
		18	-0.427530791147525\\
		19	-0.431488741232158\\
		20	-0.476578415607079\\
		21	0\\
		22	0.380251857910977\\
		23	-0.486885218893427\\
		24	-0.841235356553206\\
		25	-1.22691710714342\\
		26	-1.4724710948848\\
		27	-1.07685567968914\\
		28	-1.12598600153584\\
		29	-1.73733749471403\\
		30	-2.09051473824032\\
		31	-1.42889685771079\\
		32	-0.871740312898325\\
		33	-0.319098653813263\\
	};
% Pit stops
		\addplot [color=black, dashed, forget plot,line width=1pt]
	table[row sep=crcr]{%
		12 5\\
		12	-5\\
	};
	\addplot [color=black, dashed, forget plot,line width=1pt]
	table[row sep=crcr]{%
		21	5\\
		21 -5 \\
	};
	\end{groupplot}

\end{tikzpicture}%
\vspace{-5pt}
	\caption{Comparison between the baseline strategy and the optimal solution for the total time gap and the battery State of Charge (SoC) $\xi$, where the baseline strategy represents is always overtake. The optimal strategy saves energy by staying behind competitors and overtaking at more favorable locations on the track, resulting in a total time saving of about \unit[21.4]{s}.  }
	\label{fig:comparison1}
\end{figure}

To validate the decision making, we compare our proposed strategy with a baseline strategy that always attempts an overtake. Fig.~\ref{fig:comparison1} shows the time gap and difference in battery energy between the baseline and optimal strategy. The time gap at the end of the race is around \unit[21.4]{s}, demonstrating that optimizing both the race strategy as well as the decision making has significant benefits. By staying behind, the optimal strategy saves energy, resulting in a shorter charging time at the second pit stop. Furthermore, in the last stint after lap 30, the surplus of battery energy can be used to drive faster lap times, compared to the baseline. 

\begin{figure}
	\centering 
	\input{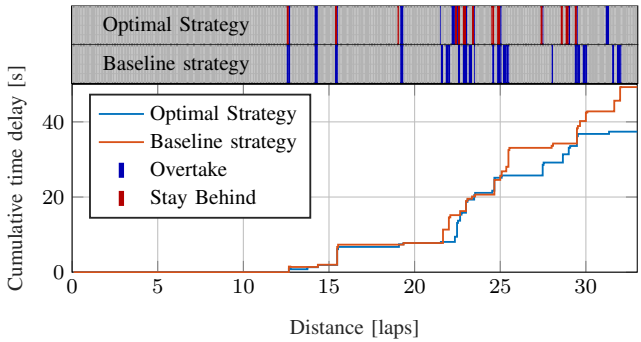}\vspace{-5pt}
	\caption{Analysis of the cumulative time delay due to competitor interactions for the baseline and optimal strategy. The timelines at the top represent the actions taken by the respective strategy. The optimal strategy shows a lower time delay compared to the baseline at the end of the race, since attempting an overtake does not guarantee a successful overtake maneuver. }\vspace{-10pt}
	\label{fig:comparison2}
\end{figure}

Lastly, we investigate the cumulative time delay due to the interactions for both strategies. Fig.~\ref{fig:comparison2} shows the interaction events for both strategies together with the cumulative time delay. We observe that the optimal strategy has a lower time delay overall, which further contributes to the time gap with the baseline. For example, at lap 25, the time gap between both strategies is almost negligible, yet the optimal strategy has more battery energy reserve. Therefore, it performs a successful overtake, while the baseline strategy performs a series of unsuccessful overtake attempts, significantly increasing the time delay. This shows that staying behind competitors instead of attempting to overtake them can be faster and extremely important to account for in the course of a race. 

%This shows that even though the time penalties are estimates partially based on heuristics, the decision making algorithm results in a significant advantage over an always overtake strategy. 

 % When attempting an overtake, we spend more energy. Overtakes themselves are still optimized.

% At end of results section / validation with always overtake decision, highlight that even though the time penalties are based on some engineering knowledge/heuristics, there is a clear advantage.

% Results are shown for a 1 hour race around the Zandvoort circuit, using data of the supercar challenge [reference to website].

\section{Conclusion} \label{Conclusion}

In this paper, we have presented a control framework to optimize the endurance race strategy for a fully electric vehicle online and accounting for competitors' interactions. To this end, we reformulated an existing race strategy optimization framework to an online approach, capturing the pit stop decision on a lap-basis. In addition, we derived time penalty functions for all major interactions and decisions to obtain the best action w.r.t.\ the competitors.
Our results showed that accounting for uncertainty in overtaking maneuvers and staying behind competitors, although possibly counter-intuitive, was \unit[21]{s} faster over the course of a \unit[1]{h} race. Ultimately, these methods bring e-race cars one step closer to \gls{acr:ice} race cars.
In future work, we plan to implement these strategies in a real-life situation on the race track.

% Contribution:
% - Lap based MPC for endurance race / online race strategy optimization
% - Decision making for overtaking, stay behind or pit / competitor interactions

% Paper outline:
% - Introduction
% 	- intro endurance racing
% 	- intro electric racing
%  	- energy management is key
% 	- lots of energy can be gained by drafting, but also lost during overtaking. Cite Braghin, Liu with slipstream effects, Duhr energy allocation etc.
% 	- So what is the optimal strategy?
% 	- Guess what we do.

% - Methods
% 	- offline maps -> lap and stint based. Take stint maps and then cut in laps?
%  	- race strategy optimization
%	- Competitor interaction

% Appendix before acknowledgment according to IEEE transactions template
\ifextendedversion
\section*{Appendix} \label{sec:Appendix}

\else

\fi

\section*{Acknowledgment}
\noindent We thank Dr.~I.~New for proofreading this paper. This paper was partly supported by the NEON research project (project number 17628 of the Crossover program which is (partly) financed by the Dutch Research Council (NWO)).

\bibliographystyle{IEEEtran}
\bibliography{Report/bibliography.bib,main,SML_papers}
%\newpage

\end{document}